# Dynamic hardware system for cascade SVM classification of melanoma

Shereen Afifi✉, Hamid GholamHosseini
*Department of Electrical and Electronic Engineering, Auckland University of Technology Auckland 1010, New Zealand*
safifi@aut.ac.nz, hgholamh@aut.ac.nz

Roopak Sinha
*Department of IT and Software Engineering Auckland University of Technology Auckland 1010, New Zealand*
rsinha@aut.ac.nz

**Abstract**

*Melanoma is the most dangerous form of skin cancer, which is responsible for the majority of skin cancer-related deaths. Early diagnosis of melanoma can significantly reduce mortality rates and treatment costs. Therefore, skin cancer specialists are using image-based diagnostic tools for detecting melanoma earlier. We aim to develop a handheld device featured with low cost and high performance to enhance early detection of melanoma at the primary healthcare. But, developing this device is very challenging due to the complicated computations required by the embedded diagnosis system. Thus, we aim to exploit the recent hardware technology in reconfigurable computing to achieve a high-performance embedded system at low cost. Support vector machine (SVM) is a common classifier that shows high accuracy for classifying melanoma within the diagnosis system and is considered as the most compute-intensive task in the system. In this paper, we propose a dynamic hardware system for implementing a cascade SVM classifier on FPGA for early melanoma detection. A multi-core architecture is proposed to implement a two-stage cascade classifier using two classifiers with accuracies of 98% and 73%. The hardware implementation results were optimized by using the dynamic partial reconfiguration technology, where very low resource utilization of 1% slices and power consumption of 1.5 W were achieved. Consequently, the implemented dynamic hardware system meets vital embedded system constraints of high performance and low cost, resource utilization, and power consumption, while achieving efficient classification with high accuracy.*

## 1. INTRODUCTION

Melanoma is the deadliest form of skin cancer. Australia and New Zealand have the highest rate of melanoma worldwide. The best cure for melanoma is early detection. Dermatologists are recently using image-based computer-aided diagnosis (CAD) systems as diagnostic tools to help them in their decision and detecting melanoma at an early stage. However, such tools are very costly and only available at the dermatologist. Therefore, a low-cost handheld device dedicated for early detection of melanoma is required to support primary healthcare providers. But, developing this device is very challenging because of the complicated computations required by the image analysis and classification algorithms within the embedded CAD system. To overcome this challenge, recent hardware advances and technologies in reconfigurable computing should be exploited to achieve cost- and energy-efficient embedded system with high performance.

Field-programmable gate array (FPGA) is a robust highly parallel processing reconfigurable device. FPGAs have been widely used for achieving required performance of embedded systems, while effectively utilizing hardware resources and achieving low power consumption at low cost [1]. For various applications, FPGAs have demonstrated significant performance and acceleration with efficient hardware implementation results, outperforming other comparable platforms such as general-purpose processors and graphics processing units [2–5]. In addition, FPGA is featured with a massive dynamic partially reconfiguration (DPR) technology that allows reconfiguring dynamically selected areas on FPGA on-the-fly, while other parts are still working (also called runtime reconfiguration) [6]. The module-based PR is widely used for various designs and applications, where reconfigurable modules (RMs) are reconfigured at runtime as depicted in Fig. 1. DPR offers design flexibility, design space expansion, power and area savings with speedups. Subsequently, FPGA is a promising hardware platform for implementing an embedded system for the CAD system, offering high-performance computing with more flexibility at low cost.

In our research group, some software development is performed for implementing different stages within the CAD system for melanoma detection (image pre-processing, segmentation, feature extraction, and classification stages) [7]. The final classification stage is remarked as the most compute-intensive stage within the CAD system that is very critical and needs hardware implementation. Based on experimental results and performance comparisons of tested classifiers, support vector machine (SVM) showed better accuracy results for classification and diagnosis of melanoma [7] (and more improvement was achieved by using the cascade SVM).

SVM classifier is a common supervised machine learning tool, which is widely used in different classification problems and applications offering efficient and high classification accuracy. Accordingly, we are focusing on implementing SVM on hardware for enhancing early detection of melanoma at low cost.

Some existing research works implemented the SVM model on FPGA using different hardware architectures/ designs aiming to achieve high performance [8]. Nevertheless, meeting vital constraints of embedded systems as high performance and low cost is very challenging, in addition to reaching an effective classification system that offers high accuracy rate. Also from reviewing the literature, it is found that using DPR feature is not efficiently and widely used, which should be exploited for improving FPGA implementations. As well, existing FPGA implementation of the special cascade SVM classifier that improves classification performance is very limited and proposed by only one research group. The cascade classification architecture consists of multi-classification stages (classifiers) designed in a cascading structure. In the cascade scheme, the majority of data are classified at the early stages that are based on simple classifiers with low complexity, leaving very little data to be classified in later stages that are more complex [9]. This motivates implementing the cascade architecture in hardware/FPGA to realize an efficient embedded classification system with high performance and low cost.

In this paper, we are proposing a dynamic hardware system for realizing an adaptive, flexible, and scalable embedded system of a cascade SVM classification targeting early detection of melanoma. An initial implementation of the cascade SVM classifier is first introduced in [10], to be extended and improved in this paper by effectively using the unique DPR technology of the FPGA aiming to optimize hardware implementation results. In addition, an intensive design description, an extensive results analysis, and a comprehensive comparative study among our previous implementations and other related works in the literature are reported. The hardware implementation results are significantly optimized with the proposed dynamic design meeting important embedded system constraints, which promises to realize a low-cost handheld medical scanning device for early detection of melanoma.

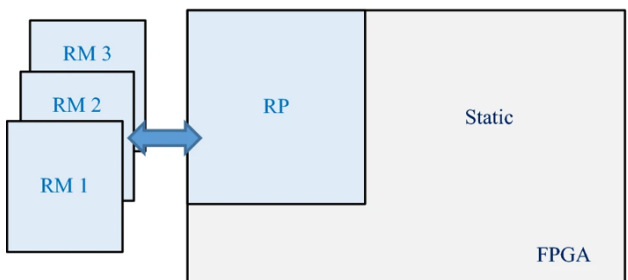

**Fig. 1** DRR architecture

## 2. RELATED WORK

Most existing research works in the literature focus on implementing the SVM classification phase online on FPGA after executing the training phase offline in software. Different FPGA-based hardware techniques have been utilized for implementing the SVM classifier [8]. Binary classifiers using linear kernel are mostly targeted for the hardware implementations. Most existing hardware designs utilize the common pipelining technique, exploiting the parallel processing capabilities of the FPGA in order to accelerate the classification task. Other hardware designs are implemented using parallel systolic array architecture and multiplier-less approach, while old versions of FPGAs are utilized for implementations without considering crucial embedded systems constrains like low power consumption constraint.

From reviewing existing work in the literature, we identify a unique research group that is exclusively working on FPGA-based cascade SVM classification architecture for object detection [9, 11–15]. Good results are achieved for real-time processing of high-resolution images in terms of fps. But, their implemented architectures suffer from slight loss in detection accuracy rates. Also, the power consumption results (greater than 3W) are considered high for embedded systems deployment. Another research group led by Hussain [16–18] is exploiting the FPGA-based DPR feature with implementing a parallel systolic array architecture, targeting classification of bioinformatics microarray data. They are also proposing a multi-core architecture of different SVM copies. But, no specific application is applied to employ the proposed architectures and so no classification accuracy is measured as well as not all implementation results are provided.

Subsequently, it is concluded that the main challenges are the difficulty of meeting important embedded system constraints of high performance, flexibility, scalability, and low levels of area, cost, and power consumption, while achieving reliable effective classification system with a high classification accuracy. Also, very limited work exists for hardware implementation of the cascade SVM classifier as well as using the powerful FPGA-based DPR feature. To the best of our knowledge, no FPGA implementation of the (cascade) SVM classifier exploiting the benefit of the hybrid architecture (hardware/software system) of the Zynq SoC exists in the literature. Additionally, almost all previous FPGA implementations are designed utilizing the traditional time-consuming hardware description language (HDL) that necessitates expert hardware designers. Nevertheless, the recent UltraFast High-Level Synthesis (HLS) design methodology is newly recommended to simplify FPGA designs [19]. Furthermore, no FPGA implementation exists in the literature for classifying melanoma clinical images using SVM. Consequently, our research work focuses on proposing a hardware implementation of an embedded SVM classification system on Zynq SoC by utilizing the HLS design methodology, targeting melanoma detection with high performance and low cost. Specifically in this paper, we are addressing the existing gap in the literature by proposing a hardware implementation of a dynamic and adaptive cascade SVM classifier exploiting the massive DPR feature of the FPGA, targeting flexibility, scalability, applicability, and adaptability while meeting critical embedded system constraints.

# 3. PROPOSED HARDWARE DESIGNS AND IMPLEMENTATIONS

This paper proposes a novel scalable architecture to implement an improved version of our proposed cascade SVM architecture that is first introduced in our published paper [10] by applying the massive DPR feature of the FPGA. In this section, the FPGA platform and development tools used are introduced. Then, the design of a single (monolithic) SVM model that is used in the cascade architecture is presented and followed by the proposed multi-core architecture (cascade architecture). Finally, the DPR-based architecture is introduced.

## 3.1 FPGA platform and system development tools

The recent FPGA platform ''Xilinx Zynq-7000 All Programmable System on Chip (SoC)'' is chosen for the SVM implementation to exploit the newest technology and reach a powerful and efficient embedded system [20]. The Zynq SoC is characterized by its hybrid architecture, which significantly simplifies the embedded system development process. The hardware programmability of an FPGA is combined as a programmable logic (PL) with an ARM Cortex-A9 dual-core as a processing system (PS) in a single SoC. Hence, the Zynq SoC is an ideal platform to develop a high-performance smart embedded system for realizing a cost-effective medical scanning handheld device.

The latest software tool ''Xilinx Vivado Design Suite'' is selected as being a powerful system design tool that simplifies embedded system design on a single device based on integrating an FPGA within a single SoC [21]. Vivado Design Suite applies the UltraFast Design Methodology that employs Design Rule Check rules, which provide guidance on HDL code and Xilinx Design Constraints in order to improve the quality of the design earlier in the flows. The Vivado tool offers a powerful intellectual property (IP) and system-centric integration with fast verification, reaching significant acceleration for design integration and implementation. So, the Vivado tool with the UltraFast Design Methodology is exploited for our embedded system development, because of its benefits of faster compile times and convergence, more predictable results, and accelerated time to market.

In addition, Xilinx Vivado Suite comprises an efficient design tool that employs the modern UltraFast High-Level Synthesis (HLS) design methodology. The HLS methodology is characterized with simplifying FPGA programming via using the high-level language replacing the traditional HDL [19]. The HLS method is highly recommended as it significantly decreases the FPGA development effort and time, while realizing an effective optimized embedded system/accelerator [22, 23]. Consequently, the Xilinx Vivado HLS tool is utilized to implement an SVM IP to be integrated into a single SoC for reaching an online embedded classification system running on the recent Zynq SoC.

## 3.2 Proposed monolithic SVM HLS IP

An algorithm of the SVM is designed to be written in C/ C ++ language using the HLS tool to implement an SVM model (to be used in the cascade system). The SVM algorithm basically implements the following decision function (1) using the linear kernel for classifying a new data sample x,

$$F(\vec{x}) = sign\left(\sum_{i=1}^{SV} \alpha_i y_i (\vec{x_i} \cdot \vec{x}) - b\right) \quad (1)$$

where a, y, and b are parameters specified from the training phase and SV denotes support vector [8]. In order to simplify the required calculations on hardware, the function is divided into two main functions (2) and (3). The first function (2) computes the summation of all SVs multiplied by the corresponding ay to be stored in an accumulated array/vector ''AC.'' Then, it is used in the second function (3) as part of the dot product with the features of the test instance x to calculate the distance value for classifying according to its sign value. The final result F(x) of the SVM classifier (decision function) has two possible values equal to 1, or − 1, which corresponds to melanoma class, or non-melanoma class, respectively.

$$\vec{AC} = \sum_{i=1}^{SV} \alpha_i y_i \vec{x_i} \quad (2)$$

$$F(\vec{x}) = sign\left((\vec{AC} \cdot \vec{x}) - b\right). \quad (3)$$

In the Vivado HLS tool using C/C ++ language, a top function module is designed as an HLS IP that basically computes the decision function (3) to be used for online classification on Zynq SoC. This hardware design depends on the number of features to implement any SVM model. The designed HLS IP stores the AC array data after it is calculated offline from a proposed software function that implements (2) on software. The proposed IP takes a 1D array input that consists of features of the test instance to produce its classified class. The proposed pseudocode of the top module function to implement function (3) is illustrated in Fig. 2.

The HLS tool provides various directives to be applied for the IP to assign different interfaces and apply other hardware and optimization techniques [24]. The AXI-lite bus is allocated as a control bus of the module in order to control the designed IP core and other connected cores in the system as well as controlling the data flow of the system through communicating with the ARM processor/PS. For optimization, the pipelining technique is applied to the for-loop using the HLS directives for enhancing data latency and throughput. By using the pipelining, the HLS synthesis

```
Define number of features+1 as F

Function: Classify ( X [F] ) {
    Set value of "b"
    Initialize "AC_array [F] " with accumulated SVs data
    FOR each feature
            Distance_value += AC_array [F] * X [F]
    END FOR
    Distance_value - = b
    IF (Distance_value ≥ 0) THEN
            RETURN 1
    ELSE
            RETURN -1
    END IF
}
```

**Fig. 2** Proposed pseudocode of the SVM HLS IP

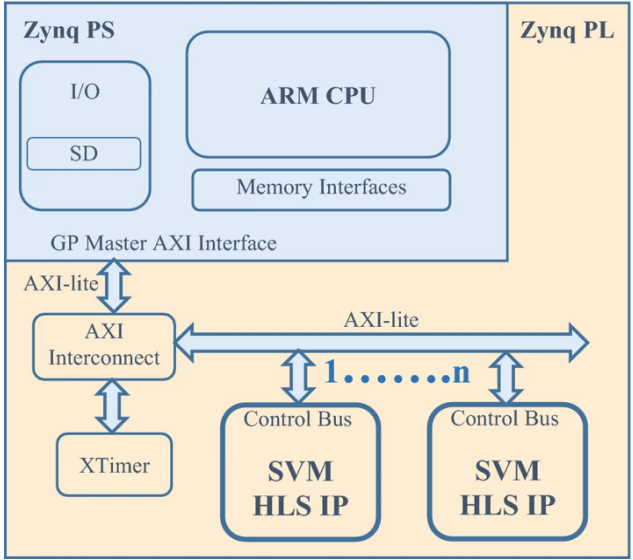

Fig. 3 Proposed hardware/software system on Zynq SoC

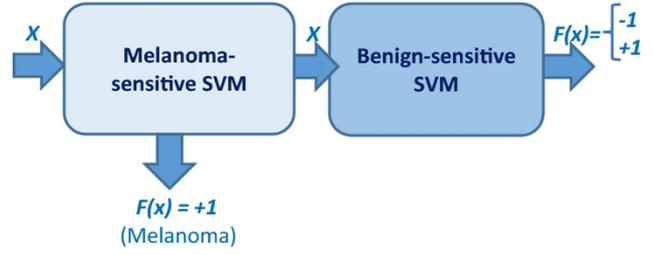

Fig. 4 Proposed two-stage cascade SVM classification system

results show latency reduction from 278 to 148 clock cycles for implementing an SVM model with 27 features. Accordingly, the speedup is nearly doubled from using the pipelining, but with some extra logic resources utilized. This founding indicates the trade-off exists between the performance and area/cost.

The proposed HLS IP of the SVM classifier is successfully co-simulated (RTL simulation) and exported as an RTL implementation (packaged IP), after synthesizing the designed code utilizing the Vivado HLS tool. Next, the exported HLS IP is integrated into the proposed hardware/software system as shown in Fig. 3. At this stage, only one IP is added ($n = 1$), to be realized on Zynq SoC. Using the Vivado Design Suite, the exported HLS IP is connected as a Zynq coprocessor in the PL part to the ARM processor in PS part through the AXI-lite interface passing by an AXI Interconnect IP. Besides, an AXI-Timer IP is exploited for performance comparisons based on the number of clock cycles needed by the IP/cores.

Finally, the designed Zynq SoC is exported for the SDK tool after successfully passing the synthesis, implementation, and bitstream generation stages in the Vivado design tool, in order to run an application on Zynq SoC. A test bench or a software program has been developed for testing and verifying the implemented monolithic SVM IP by running the implemented hardware/software system on the Zynq SoC. The ARM processor in the Zynq PS part is responsible for executing the test bench besides controlling the attached IPs/cores and the data flow in the system. A software program is designed and implemented in C using the SDK tool. The test instance is read from a text file saved on the SD card of the Zynq evaluation board. The file is read byte by byte to be parsed and then saved into an array as the main input of the implemented SVM IP on the Zynq PL. Then, the data of the array are passed by the ARM processor through the AXI-lite interface to be processed by the implemented SVM IP in order to return the classification result.

### 3.3 Proposed multi-core architecture
After implementing an SVM HLS IP, it is integrated in the proposed embedded system on Zynq SoC as in Fig. 3, where more IPs (1 $to$ $n$) can be added for design extension and scalability. By using the proposed HLS-based design, various copies of SVM IPs with different parameters/sizes could be implemented to construct a multi-core architecture that consists of n cores/IPs in a single system on chip/device. The proposed scalable multi-core architecture could be applied as an ensemble, multi-class, or cascade classifier for different classification problems and applications.

In this paper, we are proposing an SVM-based cascade classifier for improving early detection of melanoma by exploiting the proposed multi-core architecture. A dual-core architecture ($n = 2$) is proposed as our case study to be employed as a two-stage cascade SVM classification architecture. Figure 4 shows the proposed two-stage cascade classifier that consists of two SVM HLS IPs. For enhancing early melanoma detection, the first stage in the proposed cascade is a melanoma-sensitive SVM model that is trained with a dataset of more melanoma samples, while the second stage is a benign-sensitive (non-melanoma-sensitive) SVM. The scenario of this cascade architecture is proposed to verify the benign samples classification for this sensitive application, aiming to decrease the risk of false detection by reducing the number of false negative. The test instance X is passed first by the early stage to detect melanoma as a positive sample; otherwise, it is passed to the second stage for a second verification.

An application is also developed in the SDK tool (similar to the proposed test bench for the monolithic SVM) to validate the implemented two-stage cascade classifier on Zynq SoC. Different scenarios could be easily applied to the proposed cascade architecture handling different classification problems and requirements. Also, the proposed two-stage cascade classifier would be extended in the future to a multistages cascade classifier by simply adding more IPs, targeting performance improvement of classification.

### 3.4 Proposed dynamic cascade SVM architecture
The FPGA-based DPR technique is exploited to achieve a dynamic hardware system for realizing an adaptive SVM cascade classifier on Zynq SoC, targeting more flexibility and scalability in addition to more improvements in area, power, and classification performance. In order to apply the DPR feature on the proposed two-stage cascade architecture, a static top-level design with a single reconfigurable partition (RP) module/block is first proposed. The proposed system in Fig. 3 with one SVM HLS IP/core ($n = 1$) is used to define the static design wrapper, while the SVM HLS IP is assigned as a RP to be then defined as a black box (BB) module. The BB module is then used to be reconfigured dynamically (swapped in and out) at runtime to one of the two implemented SVM IPs, melanoma- and benign-sensitive IPs. Each IP is defined as a reconfigurable module (RM) for the RP that allocates the logic size based on the larger resource utilization. Then, two different configurations are defined that combines the static design wrapper with each RM. The first configuration is for instantiating the first stage/IP, while the second configuration instantiates the second stage in the cascade.

Each of the three defined configurations (static with BB/RM) are synthesized and implemented using the Vivado Design Suite, where the floorplanning tool is used to design the floorplan of the RP in the device with the associated resources. Finally, two bitstreams are generated for each configuration, full and partial bitstreams to form a library of configuration files for configuring the device/FPGA. The full bitstream should be used for the first configuration of the FPGA to configure the whole device. Then, the partial bitstream is used to dynamically reconfigure the RP only to the corresponding RM, which has much shorter configuration time than the full bitstream.

For testing the implemented dynamic cascade classifier on Zynq SoC, the developed application in the SDK tool is modified to include dynamically reconfiguration using the generated partial bitstreams. Two configuration modes can be used for partial reconfiguration by using either JTAG or PCAP (processor configuration access port) interface. The JTAG interface is used in our experiments, while the PCAP is to be used in the future to allow reconfiguration through the application running on the ARM processor, aiming for more time improvement.

The proposed DPR-based dynamic hardware system could be easily extended in the future to include more RMs reconfiguration for realizing an adaptive multistages cascaded classifier, while keeping same design space. Therefore, our proposed dynamic hardware system promises to meet challenging embedded system constraints for realizing a low-cost handheld device for early detection of melanoma.

## 4. EXPERIMENTAL RESULTS AND DISCUSSION

The modern Xilinx Vivado 2016.1 Design Suite is utilized to design and implement our proposed system on the Xilinx XC7Z020CLG484-1 target device of the Zynq-7 ZC702 Evaluation Board. The Vivado HLS tool is used first to design and implement each SVM HLS IP. Then, the developed SVM IP is exported to be integrated into the proposed Zynq SoC (Fig. 3) that is designed in a block diagram using the Vivado tool. Each of the designed Zynq systems is synthesized, placed, and routed, and finally, the bitstream is generated to be exported for the Xilinx SDK tool to run an online classification application on Zynq for testing and verification. For the DPR-based hardware system, three configurations are implemented as a static top-level design with BB module, RM-M (for melanoma-sensitive SVM) and RM-N (for non-melanoma-sensitive SVM) by the aid of the Vivado tool using the floorplanning tool. For each configuration, two bitstreams are generated, full and partial bitstreams for different configuration options.

In the next subsections, the implemented SVM models are introduced, and then different experimental results and comparisons are presented and discussed.

### 4.1 Implemented SVM models

A common SVM classifier called ''SVM-Light'' is studied as a case study to implement our SVM IP for melanoma detection. The SVM-Light is a robust and simple classifier, which is implemented in C and is widely used in various classification problems [25]. The modern UltraFast HLS design methodology is utilized to design and implement an SVM IP on the Zynq SoC, which is based on the binary classification algorithm and C/C ++ code of the SVM-Light. In the designed C/C ++ code, float data type is assigned for all used data and is mapped to the standard single-precision floating-point format in the hardware.

The training phase is done offline on software by exploiting the available SVM-Light windows application, where the default parameters and the linear kernel function are used to generate the trained SVM models. Based on our previous work within our research group for melanoma detection [7], a dataset is used for model training that consists of a total of 356 clinical images, including 168 melanoma and 188 benign images. In order to form a features dataset for the training of the SVM model, some selected pre-processing, segmentation and feature extraction (based on HSV color channels) algorithms are applied to the images dataset (512 × 512 pixels) (SoC implementation of these used algorithms will be investigated in the future for realizing the targeted handheld device). Finally, a new dataset of 356 instances of 27 features each is extracted from the images dataset, to be used for training SVM models [7]. This dataset is manipulated to build two different datasets in order to generate a melanoma-sensitive model as model M and non-melanoma-sensitive (benign) model as model N to construct the proposed two-stage cascade architecture. In order to achieve a higher accuracy for the trained models, data scaling and normalization techniques are applied to the original dataset as well as using the cross-validation technique in the training phase. Table 1 summarizes different parameters of the two implemented SVM models. For each model, the table shows the number of SVs generated with the number of instances used in the training dataset of 27 features each. Finally after generating the trained SVM models offline, the models' data are extracted to implement the trained SVM model on hardware using this proposed design that depends on the number of features.

### 4.2 Classification accuracy

Using the Xilinx SDK tool, an application of classifying a test instance is developed in C to test and validate the implemented classification systems. Some test instances of extracted features are correctly classified by all implemented systems (one at a time). Regarding the classification accuracy, the two models M and N are trained using the cross-validation method to produce good accuracy of 97.92 and 72.51%, respectively. Both models are used in the implemented cascade scheme for improving classification performance and verifying melanoma diagnosis. The experimental results demonstrated that every hardware classification result of the implemented classifiers is exactly equal to the corresponding software classification result. In order to verify the hardware classification result of the implemented SVM IP and compare it with the software result, we monitored the calculated distance value in the decision function (3) for the test instance before applying the sign function ("Distance_value" in pseudocode in Fig. 2) for determining its class ($+1\ or\ -1$). By using the C/RTL co-simulation results from the HLS tool, the distance value from all implemented IPs is easily compared to that value generated from the SVM-Light window application/software to be identical for all tested instances. Accordingly, the percentage error is equal to zero, preserving the classification accuracy level without any loss from the hardware implementation, in contrast to some existing implementations in the literature as stated in Sect. 2. More instances are to be tested in the future aiming to validate the online classification accuracy of the hardware implemented classifiers on Zynq SoC, in addition to calculating sensitivity and specificity metrics.

Table 1 Parameters of implemented SVM models

| SVM model | Training dataset | | SVs | Accuracy % |
|---|---|---|---|---|
| | Melanoma | Benign | | |
| Model M | 100 | 44 | 61 | 97.92 |
| Model N | 67 | 144 | 139 | 72.51 |

**4.3 Processing time and speed**
From the HLS tool, the synthesis results show processing time of 1.5 ls that is required for each model M and N using frequency of 100 MHz. For the two-stage cascade architecture, the processing time is nearly doubled (3 $\mu$s) if the test sample is negative (non-melanoma) and is equal to one IP's processing time otherwise. For verifying this synthesis/simulation results, the XTimer IP is connected to the designed SoC as in Fig. 3 to measure clock cycles required for running the IP on Zynq PL through executing the developed software program (application) on the PS ARM processor using the Xilinx SDK tool. For 250 MHz, less processing time of 1.8 ls is demonstrated for the cascade architecture [10], and a speedup of 5 × is reached compared to a software implementation running on ARM processor.

**4.4 Hardware resource utilization and power consumption**
Table 2 summarizes the implementation results of all implemented systems/configurations on Zynq SoC using 100 MHz frequency. Overall, all resource utilization percentages are considered significantly low while dissipating low power of only 1.5 W for all implemented systems, which promises more design extension and expansion for larger-scale problems by using our proposed design as well as meeting critical embedded system constraints. The power consumption results are reported by Vivado tool (the confidence level is medium). The device statically dissipates 10% of the total power consumption, whereas the remainder 90% is consumed by the dynamic activity, mostly by the Zynq PS component (95% of total dynamic power) compared to other on-chip components. The implemented two-stage cascade architecture utilizes almost double of the monolithic-based system utilization for all hardware resources and with only two extra memory LUTs, while power consumption increases by only 0.2 W. The two monolithic models M and N utilize equal resources and power as both have the same size of 27 features. For the DPR cascade system, both RM-M and RM-N configurations utilize equal resources and power to be dynamically reconfigured at runtime, which is less than the baseline cascade architecture and nearly equal utilization to the monolithic model (with slightly extra slices and LUTs) with only 0.1 excess power dissipation. That shows that by using our novel DPR-based design, we can realize the cascade scheme with efficient resources utilization and power consumption that is nearly equal to a single SVM model, while gaining advantages of using the cascade scheme of improving classification speed (for positive/melanoma samples) and accuracy as well as diagnosis verification.

**4.5 Comparison with our previous proposed designs**
Three papers are published regarding hardware implementation of a binary SVM classifier with linear kernel for melanoma detection on the recent hybrid Zynq SoC using the latest UltraFast HLS design methodology. Our initial hardware/software co-design [26] was proposed in order to implement the complicated dot-product calculation onto the hardware/PL (similar to this proposal), and the test data are streamed using a stream interface passing by a DMA IP. Then, the hardware design was extended to implement the whole function (full SVM) on the PL as a Zynq coprocessor/accelerator in [27], which depends on both numbers of SVs and features. Another similar design was proposed [10], which used BRAM interfaces to pass required data instead of using the stream interface with the DMA IP. This proposed design simplifies the previous SVM IP designs to achieve lesser area, power, and cost for design extension and realizing multi-core (cascade) classification architecture, aiming to improve the classification performance.

Table 3 shows a comparison of implementation results between our previous designs and this proposal aiming to find an optimum solution to balance the existing trade-off between performance and cost (area and power). The implementation results (Table 1) in this proposal are updated to include an AXI-Timer IP in the implemented system in order to provide a fair comparison with our previous implemented systems that consist of the same timer IP (extra slices and LUTs are utilized). The number of features is constant for all implemented models, equal to 27 features, while other parameters are different among them. Different optimization techniques are applied to all proposed designs like loop unrolling and pipelining for optimizing the implementation results. In addition, the implemented hardware SVM systems showed speedup on hardware (Zynq PL) over software implementations on embedded ARM processor (Zynq PS).

For the monolithic SVM, less resources are utilized with only one BRAM and same number of DSPs (5) using this proposal. In addition, the least power consumption of only

**Table 2** Implementation results on Zynq SoC

| SVM model | Resources utilization (%) | | | | | P (W) |
|---|---|---|---|---|---|---|
| | Slices (106,400) | LUT (53,200) | LUT-RAM (17,400) | BRAM (140) | DSP (220) | |
| Model M | 1046 | 858 | 70 | 1 | 5 | 1.54 |
| | (1%) | (1.6%) | (0.4%) | (0.7%) | (2.3%) | |
| Model M | 1046 | 856 | 70 | 1 | 5 | 1.54 |
| | (1%) | (1.6%) | (0.4%) | (0.7%) | (2.3%) | |
| Cascaded model | 1785 | 1478 | 72 | 2 | 10 | 1.56 |
| | (1.7%) | (2.8%) | (0.4%) | (1.4%) | (4.6%) | |
| RM-M configuration | 1050 | 867 | 70 | 1 | 5 | 1.55 |
| | (1%) | (1.6%) | (0.4%) | (0.7%) | (2.3%) | |
| RM-N configuration | 1050 | 862 | 70 | 1 | 5 | 1.55 |
| | (1%) | (1.6%) | (0.4%) | (0.7%) | (2.3%) | |

**Table 3** Implementation results comparison

| Results | Monolithic SVM model | | | | | | Cascaded SVM | | |
|---|---|---|---|---|---|---|---|---|---|
| | [26] | [27] | [10] 1 | [10] 2 | Proposed M | Proposed N | [10] | Proposed | Proposed DPR |
| Resources utilization | | | | | | | | | |
| Slices | 5584 | 2676 | 10,874 | 30,006 | **1306** | **1302** | 4304 | **2001** | **1310** |
| LUTs | 4373 | 2267 | 7218 | 17,506 | **1448** | **1448** | 3414 | **1762** | **1457** |
| Memory LUT | 173 | 204 | 874 | 2873 | **70** | **70** | 215 | **72** | **70** |
| BRAM | 3 | 12 | 48 | 48 | **1** | **1** | 2 | **2** | **1** |
| DSP48 | 5 | 5 | 5 | 5 | **5** | **5** | 10 | **10** | **5** |
| Power (W) | 1.74 | 1.75 | 2.06 | 2.65 | **1.54** | **1.54** | 1.74 | **1.56** | **1.55** |
| Frequency (MHz) | 50 | 100 | 250 | 250 | **100** | **100** | 250 | **100** | **100** |
| Processing time (μs) | 3.4 | 141.4 | 11.46 | 39.3 | **1.5** | **1.5** | 1.8 | **3** | **3** |
| SVs# | 346 | 248 | 61 | 248 | **61** | **139** | 61 + 139 | **61 + 139** | **61/139** |

Bold values are those of the proposed design

1.5W is achieved. Processing time is significantly decreased to 1.5 ls at 100 MHz for implementing a moderate-size model with 27 features, (regardless the number of SVs) with a high classification accuracy of 97.9%. For the two-stage cascade classifier, less resources and power are achieved by using less frequency, while processing time is slightly increased. That shows the trade-off of extra cost required for gaining higher speed. By using the DPR feature, both resources utilization and power dissipation are significantly reduced at the cost of the configuration time (that is reduced by using the partial bitstream for reconfiguration). Accordingly with the proposed dynamic design, more design extension and scalability could be easily achieved at low cost, area, and power, while getting use of advantages of multi-stages cascade architecture.

**4.6 Comparison with related works**

Some existing related FPGA implementations of binary SVMs with different kernel types for various applications are selected to be compared with our implemented Zynq systems of both monolithic and cascade SVM classifiers that are summarized in Table 4a and b, respectively. Using the recent UltraFast HLS design methodology, our hardware implementations on the recent hybrid Zynq SoC achieved significant hardware results compared to others that used old versions of FPGAs for implementing the traditional pipelined designs, multiplier-less method, and common systolic array architectures. It is clear that the least resource utilization and power consumption are demonstrated by our implemented systems with real moderate size of SVM parameters applying the linear kernel. The 1.5 W of power is significantly low compared to a very high power of 15 Win[4] and high power of 3 W in [9], while others did not consider this critical embedded system constraint. In addition, the least processing time of only 1.5 ls is achieved at 100 MHz operating frequency that is extremely less than others, while 0.9 ls [10] is demonstrated using 250 MHz that is equal to [28] using a higher frequency of 370 MHz. Moreover, the highest classification accuracy level of almost 98% is demonstrated, while some did not verify their hardware classifiers [4, 17, 29]. Also, our proposal preserved the classification accuracy level with zero loss from the software implementation compared to [9] who suffered from slightly loss in accuracy. Accordingly, our implemented models on the recent hybrid Zynq SoC platform achieved optimized results for the hardware resource utilization, power consumption, detection speed, and processing time with high classification accuracy rates using real data for melanoma detection.

Regarding the cascade architecture, only one research group exists who worked with implementing the cascade SVM architecture on FPGA, where only one paper [12] exists that used the DPR on a two-stage cascade architecture, but no implementation results are given for hardware results and power consumption. So, another implemented work [9] that uses four-stage cascade architecture and without applying the DPR is selected for this comparison with our implemented two-stage cascade architecture. Despite using less number of features and SVs for our application ''melanoma detection,'' we used for the two stages extremely less resources and power compared to the four stages implemented with higher number of SVs and dimension for face detection [9]. Besides, we demonstrated zero loss in classification accuracy, while they suffered from 0.7% reduction in accuracy. Furthermore by using the DPR, more optimization in area and power is achieved, while keeping the same accuracy level and performance with extra cost of the milliseconds of the configuration time. However, the configuration time could be optimized by using the partial bitstream than using the full bitstream.

Compared to Hussain et al. [17] who applied the DPR feature to their proposed architecture, we validated our proposed multi-core architecture to employ the cascade classification for melanoma detection using real dataset while achieving lower resources. However, they did not apply any real application to their proposal and no results are provided for power, resources of the multi-core architecture (with or without DPR), and the classification accuracy.

Finally, to the best of our knowledge, our dynamic hardware system implemented on Zynq SoC using the HLS methodology is considered to be the first FPGA-based cascade SVM classifier existing in the literature that targets melanoma classification. In addition, our implemented system successfully overcomes most challenges existing in the literature of meeting critical embedded system constraints of high performance, flexibility, scalability, and low levels of area, cost, and power consumption while reaching reliable effective classification system with a high classification accuracy.

## 5. CONCLUSIONS

A novel dynamic (DPR-based) hardware system is implemented for a cascade SVM classifier on the recent hybrid Zynq SoC using the latest UltraFast HLS design methodology targeting early detection of melanoma with high performance and low cost. A scalable multi-core architecture based on HLS SVM IP is proposed and validated with a two-stage cascade classifier implementation based on a melanoma-sensitive SVM classifier with 98% accuracy and a benign-sensitive classifier with 73% accuracy. Using the massive DPR feature of the FPGA, more optimization in implementation results is achieved, while gaining flexibility, adaptability, scalability, and applicability. Compared to reported implementations in the literature, we achieved significantly less resource utilization (1% slices) and power consumption (1.5 W) with high performance at low cost. Consequently, the implemented system on Zynq SoC meets crucial embedded system constraints, while achieving efficient and adaptive classification with high accuracy.

The implemented cascade architecture can be extended to realize multistages cascade by simply adding more SVM IPs in a single device/SoC, aiming to improve classification performance and diagnosis verification. The proposed multi-core architecture could be applied in the future as a multi-class classifier, an ensemble classifier, or different scenarios. Moreover, the DPR feature could be employed to the proposed multi-core architecture to reconfigure each core (RP) in the architecture with multi-RMs of various SVM copies or different classifiers, forming an adaptive and flexible system on a single SoC. Finally, the implemented classifier is feasible to be embedded in the future within a low-cost handheld medical scanning device for early melanoma detection.

**Compliance with ethical standards**

**Conflict of interest** The authors declare that they have no conflict of interest.

Table 4 Implementation results comparison with related works

| References | FPGA resources | | | | Power (W) | FPGA | #SVs | Dim | Accuracy % | Processing time/speed | Frequency (MHz) | Hardware design | Kernel | Application |
|---|---|---|---|---|---|---|---|---|---|---|---|---|---|---|
| | Slices | LUTs | BRAM | DSP | | | | | | | | | | |
| *(A) Monolithic SVM* | | | | | | | | | | | | | | |
| [28] | 58,688 | – | 800 | 768 | – | Virtex-6 | 760 | 128 | – | 0.9 μs | 370,096 | Pipelined | RBF | – |
| [4] | 59,208 | 122,637 | 2049 | – | 15 | Virtex-5 | 16 | – | – | 0.02 s | 200 | Pipelined | Gaussian | Skin classification |
| [29] | 12,674 | 41,135 | 132 | 64 | – | ML505 | 2048 | 2048 | – | 712.66 μs | 92 | Pipelined | Linear | Pedestrian detection |
| [30] | 23,220 | 8887 | 74 | 64 | – | ML505 | 818 | 400 | 76–78 | 40–122 fps | 100 | Systolic array | RBF Polynomial | Object detection |
| [31] | 5162 | 8887 | 74 | 64 | – | ML505 | 818 | 400 | 88 | 54 μs | 100 | Systolic array | Polynomial | Object detection |
| [17] | 1810 | 1705 | 21 | 21 | – | ML 403 | 1024 | 20 | – | 7.34 μs | 142.9 | DPR, Systolic array | Linear | Classifying biomedical data |
| Proposed M | 1046 | 858 | 1 | 5 | 1.54 | Zynq-7 | 61 | 27 | 97.9 | 1.5 μs | 100 | Pipelined HLS-based | Linear | Melanoma detection |
| *(B) Cascade SVM* | | | | | | | | | | | | | | |
| [9] four-stage | 13,038 | 31,854 | 131 | 59 | 3.2 | ML505 | 254 | 400 | 84 | 70 fps | 84 | Multiplier-less | Linear polynomial | Face detection |
| Proposed two-stage | 1785 | 1478 | 2 | 10 | 1.56 | Zynq-7 | 200 | 27 | 97.9 and 72.5 | 3 μs | 100 | Pipelined HLS-based | Linear | Melanoma detection |
| Proposed DPR | 1050 | 867 | 1 | 5 | 1.55 | Zynq-7 | 61/139 | 27 | 97.9 and 72.5 | 3 μs | 100 | DPR, Pipelined HLS-based | Linear | Melanoma detection |